\documentstyle[12pt,aasms4]{article}

\received{21 October 1996}
\accepted{24 January 1997}

\begin{document}

\title{A Statistical Comparison of Cluster Mass 
       Estimates from Optical/X-ray Observations 
       and Gravitational Lensing}

\author{Xiang-Ping Wu}
\affil{Beijing Astronomical Observatory, Chinese Academy of Sciences,
       Beijing 100080, China}

\and

\author{Li-Zhi Fang}
\affil{Department of Physics, University of Arizona, Tucson, AZ 85721}

\begin{abstract}
We present a statistical comparison of three different estimates of
cluster mass, namely, the dynamical masses obtained from the velocity
dispersion of optical galaxies, the X-ray masses measured from the
temperature of X-ray emitting gas under the assumption of isothermal
hydrostatic equilibrium, and the gravitational lensing masses derived
from the strong/weak distortions of background galaxy images. Using a
sample of 29 lensing clusters available in literature, we have shown
that the dynamical masses are in agreement with the 
gravitational lensing masses, while the X-ray method has systematically
underestimated cluster masses by a factor 2-3 as compared with the others.
These results imply that galaxies indeed trace the gravitational potential
of their clusters, and there is no bias between the velocities of the dark
matter particles and the galaxies in clusters.  The X-ray cluster mass 
discrepancy is probably from the simplification in the models for
the X-ray gas distribution and dynamical evolution.
\end{abstract}


\section{Introduction}

Clusters of galaxies are the largest coherent and gravitationally 
bounded objects in the universe. The precise determination of
their gravitational masses is crucial for our understanding of 
formation and evolution of cosmic structures,  
for mapping of matter distribution on large-scales and also 
for measurement of the present mean mass density of the universe 
($\Omega_0$). Historically, cluster masses are derived 
from the dynamical analysis of the observed velocity dispersion of
cluster galaxies based on the virial theorem, which results in 
the so-called virial cluster masses $M_v$. With the development 
of the X-ray astronomical techniques clusters can be selected 
from the X-ray emission of the hot diffuse intracluster gas,
giving rise to the X-ray cluster masses $M_x$ when combined with
the assumption of hydrostatic equilibrium. 
Over the past decade the detection of
the gravitationally distorted images of faint distant galaxies 
behind some clusters of galaxies provides another independent
mass estimate: the gravitational cluster masses $M_{lens}$.
In particular, $M_{lens}$ are obtained regardless of the cluster
matter state and components.
These three methods should be incorporate, and comparisons
of their results would yield a very useful
clue regarding the dynamical evolution of clusters and a test 
for the accuracy of cluster mass determinations.

The early studies based on a few selected clusters in which both 
lensing and optical and/or X-ray data are available claimed a 
cluster mass discrepancy by a factor of typically $2\sim3$ 
among the three methods
(Wu, 1994; Fahlman et al. 1994; Miralda-Escud\'e \& Loeb 1995),
while a consistency between dynamical masses and gravitational
lensing derived ones has been also reported in
some cases, e.g. PKS0745-191 (Allen, Fabian \& Kneib 1996).
We have recently carried out a statistical comparison of 
the overall cluster radial matter distributions determined from
X-ray observation and gravitational lensing
and concluded that the X-ray analysis may have systematically 
underestimated cluster masses  at least in the 
central regions (Wu \& Fang 1996; hereafter Paper I).

From the theoretical point of view, the above three methods have to
give the same cluster masses if clusters are dynamically
relaxed. This probably accounts for the result in clusters
like PKS0745-191 which appear to be regular in optical/X-ray
morphologies. Though one may attribute the reported mass discrepancy to 
the nonthermal pressure (Loeb \& Mao 1994; Ensslin et al. 1997) or
the projection effect (Miralda-Escud\'e \& Babul 1995;
Cen 1996), it is
most likely that the problem is relevant to the cluster matter
distribution and dynamical evolution. 
Indeed, both optical and X-ray observations
have revealed the presence of substructures in most of clusters, 
implying that clusters may be still in the era of 
formation. In particular, the recent spatially-resolved measurements of
gas temperature in some clusters  illustrate the
complex two-dimensional patterns including the asymmetric variations and
the significant decline with radius (e.g. Henriksen \& White 1996;
Henriksen \& Markevitch 1996; Markevitch 1996), 
which are the strong indicators of the effect of substructure merging. 
For those clusters,
we are unable to apply the equation of
hydrostatic equilibrium to the X-ray emitting gas. 
Therefore, it is naturally expected that
the X-ray cluster mass obtained under the assumption of 
a hydrostatic equilibrium is in principle not representative of 
the true cluster mass, and thereby should be different from 
the gravitational lensing derived mass and/or even the virial mass.  
This scenario is supported by the recent numerical study of X-ray and lensing
properties of clusters of galaxies from the standard CDM simulations
by Bartelmann \& Steinmetz (1996). They found that the cluster
masses can be biased low if the traditional $\beta$ model is 
adopted for the intracluster gas, which is especially true when
clusters exhibit pronounced substructures, while the strong
lensing preferentially selects the clusters that are dynamically
more active than the average.

Alternatively, it should be noticed
that $M_v$ derived from the virial theorem and $M_x$ derived from
the hydrostatic equilibrium have very different physical implications. 
$M_v$ is related to the galaxy velocity dispersion and galaxy number
density. Comparison of $M_v$ and $M_{lens}$ would set constraints on
the bias parameter between the velocities of the galaxies and of the dark 
matter and test whether optical galaxies trace the gravitational
potential of the cluster. On the other hand, $M_x$ depends on the
temperature variation and the density profile of the
hot diffuse gas, which may suffer from the influence of 
the possible existence of the turbulence and magnetic field.  
As a result, comparison of $M_x$ and $M_v$ may allow one to 
determine how significant the nonthermal pressure contributes 
to the computation of $M_x$ because galaxies are unaffected by 
the nonthermal pressure in clusters. Furthermore, the comparison
of the three cluster estimates may help to solve the puzzle of
the baryonic matter excess in clusters if clusters provide
a fair sample of the universal baryon fraction (see Paper I 
for summary).  Recall that the baryon fractions in clusters
are computed using the dynamical masses
$M_x$ and/or $M_v$, which may have large uncertainties if 
most of clusters are still in the process of violent merging.  
As a whole, galaxies and gas particles 
have probably experienced very different evolutions in the
formation of clusters and then exhibit different 
dynamical states and density distributions in clusters today.

Numerous lensing and optical/X-ray observations
of clusters have now made it timely and
possible to carry out these comparisons. Unlike Paper I that 
chose X-ray and lensing data separately from literature, 
we now work with the lensing clusters only, 
in which the strongly and/or weakly distorted images of background 
galaxies have been observed. While this paper was in the 
refereeing stage, we received a preprint by Smail et al. (1997)
who made a similar but sophisticated investigation 
for 11 distant clusters observed with {\it HST}. 
Throughout this paper  we adopt
a matter-dominated flat cosmological model of $\Omega_0=1$ and
a Hubble constant of $H_0=50$ km s$^{-1}$ Mpc.

\section{Cluster sample and mass estimates}

Strongly/weakly distorted images of background galaxies have been so far 
detected in $\sim40$ clusters of galaxies (Fort \& Mellier 1994;
Wu 1996; Van Waerbeke \& Mellier 1996). For our purpose we
only select those lensing clusters whose X-ray  
luminosity($L_x$)/temperature($T$)
and/or optical galaxy velocity dispersion ($\sigma$)
are available in literature. This results in a non-exhaustive 
list of lensing cluster sample (Table 1) containing
29 clusters and 39 measurements. 

\placetable{table-1}

Temperature of the hot diffuse gas is the most important parameter 
for the X-ray cluster mass determination. Unfortunately, only few clusters
in Table 1 have the measured temperature: A370 ($T=8.8\pm0.8$ keV), 
A1689 ($T=10.1_{-2.8}^{+5.4}$ keV),
A2163 ($T=13.9$ keV), A2218 ($T=6.7$ keV), MS0451($T=10.4\pm1.2$ keV) and
PKS0745 ($T=8.6_{-0.9}^{+1.1}$ keV). This arises from the difficulty of
the X-ray spectroscopic observations. Nonetheless, a correlation between
cluster temperature and bolometric X-ray luminosity $L_{x,bolo}$ 
has been well established (e.g. Edge \& Stewart 1991; Fabian et al. 1994):
$L_{x,bolo}=10^{43.06\pm0.08} T^{2.68\pm0.10}$
while $L_x$ in unit of erg s$^{-1}$ can be relatively easily obtained.  
Such a relationship thus enables us to translate $L_x$ into $T$.
Assuming a mechanism of free-free bremsstrahlung for the gas-emission
and adopting an approximate Gaunt factor given by  Mewe et al. (1986),
we have computed the cluster temperature $T$ for each cluster in Table 1.
Note that the error bars in the resulting $T$ 
account for both the uncertainties of the $L_x$
measurement and of the $L_x$--$T$ relation. It is seen that 
the agreement between the measured temperature  and the 
estimated one is fairly good for the above six clusters.

Cluster optical and X-ray morphologies reflect their
dynamical state at different evolutionary stages, 
which turn to be quite varied 
observationally. Apparently, a well-relaxed system would appear to be
regular. MS0440 and PKS0745 in Table 1 might be two good examples
of this kind.
On the other hand, clusters that are currently forming may have
very irregular morphologies.  
Substructures in optical/X-ray images are the most common feature in those
clusters and on-going collisions of the substructures are the most 
convincing evidence for interaction. 
Another indicator for the cluster non-hydrostatic process 
is the complex temperature patterns observed
recently from a number of spatially resolved measurements of the gas
temperature. All these sorts of clusters are marked by ``I'' in Table 1. 
Moreover, we utilize ``E'' to denote the rest clusters in which
substructures do not clearly present but their X-ray/optical maps
look more or less like ellipses in shape.

Fitting the azimuthally averaged X-ray surface brightness profile
by the usual $\beta$ model and assuming a hydrostatic equilibrium
for the X-ray emitting gas,  we can obtain the projected X-ray cluster mass
$m_x$ within cluster radius $r$ through (Wu 1994)
\begin{equation}
m_x=1.13\times10^{14}M_{\odot}\beta_{fit}\left(\frac{r_{xc}}{{\rm Mpc}}\right)
       \left(\frac{kT}{{\rm keV}}\right)\tilde{m}(r),
\end{equation}
where 
$$
\tilde{m}(r)=\frac{(R/r_{xc})^3}{(R/r_{xc})^2+1}-
             \int_{r/r_{xc}}^{R/r_{xc}}x\sqrt{x^2-(r/r_{xc})^2}
             \frac{3+x^2}{(1+x^2)^2}dx
$$
and $r_{xc}$ and $R$ are the core radius in $\beta$ model and the cluster 
physical radius, respectively. A straightforward computation shows 
that $\tilde{m}$ depends very weakly on $R$
and we will thus take a value of $R=3$ Mpc in the following calculation.

We now give another way to estimate the cluster mass. We 
model the cluster matter distribution by an isothermal
sphere which is characterize by 
a core radius $r_{dc}$ and the optical galaxy velocity 
dispersion $\sigma$. We use the term ``virial'' or ``dynamical'' 
cluster mass to 
denote the mass given by the optical galaxy velocity dispersion,
though this differs from the usual virial analysis which utilizes the galaxy 
number density in cluster rather than the dark matter profile. 
Our attempt here is to examine whether
$\sigma$ can provide a good mass estimate  when compared with 
gravitational lensing method. 
Yet, the radial galaxy distribution in 
cluster is not well constrained. Recall that the traditional 
King model has been questioned (Bahcall \& Lubin 1994).
For a softened isothermal sphere,
the projected ``virial'' mass within a radius of $r$ is simply
\begin{equation}
m_v=\frac{\pi\sigma^2}{G}r
\left(\sqrt{1+\frac{r_{dc}^2}{r^2}}-\frac{r_{dc}}{r}\right).
\end{equation}

Finally, with gravitational lensing method one is able to determine
the projected gravitational cluster mass within the arc position 
$r_{arc}$ or the distance $r$ from the cluster center on the cluster plane:
\begin{equation}
m_{lens}=\left\{
\begin{array}{ll}
\pi r_{arc}^2\Sigma_{crit}, \;\;\;\;\ & {\rm for\; arc/arclet};\\
\pi r^2\zeta(r) \Sigma_{crit}, \;\;\;\; & {\rm for\; weak \;lensing},\\
\end{array} \right.
\end{equation}
where $\Sigma_{crit}=(c^2/4\pi G)(D_s/D_dD_{ds})$ is the critical
surface mass density with $D_d$, $D_s$ and $D_{ds}$ being the
angular diameter distances to the cluster, to the background galaxy and 
from the cluster to the galaxy, respectively,  and
$\zeta(r)$ measures the statistics of the shear field $\gamma$ 
of background galaxy images induced by the cluster (Fahlman et al. 1994). 
Note that in eq.(3) arclike images have been presumed to trace the Einstein
radius of the cluster 
so that the alignment parameter is approximately taken to be zero. 
Furthermore, we need to make two remarks: (1)Redshift data are still not
available for nearly half of the arcs listed in Table 1, for which we
have assumed a redshift of $z_s=0.8$. This leads to the decrease of  
mass estimate $m_{lens}$ by a factor of 1.4 for a typical arc-cluster
at $z_d=0.3$ if the background galaxy is set to be $z_s=2$. 
(2)Weaking lensing 
analysis provides only a low bound on $m_{lens}$.

The resulting three mass estimates have been given in Table 1.
Uncertainties in most 
of the $m_{lens}$ results and some of the $m_v$ results are hard to 
evaluate at present. Alternatively, 
in the computation of $m_x$ we adopt a mean  core radius of 
$r_{xc}=0.25$ Mpc but allow $r_{xc}$ to vary from $0.1$ Mpc to $0.5$ Mpc. 
So, the error bars in the final result of $m_x$ include both the 
uncertainties in $T$ and the ones in $r_{xc}$. Moreover, all the
results of $m_x$ in Table 1 correspond to 
$\beta_{fit}=1$ while the observationally fitted value is around 
$\beta_{fit}\approx0.67$ (e.g. Jones \& Forman 1984). For the 
dynamical cluster mass $m_v$ given by the optical galaxy 
velocity dispersion, we have assumed  $r_{dc}=0$.

\section{Results and discussion}

Comparisons of cluster mass estimates  from  X-ray gas hydrostatic 
equilibrium, dynamical analysis and gravitational lensing are 
shown in Figure 1(a) and (b) using the data of Table 1. An immediate 
conclusion is that $m_{lens}$ agrees essentially with $m_v$ while
a systematic excess of $m_{lens}$ with respect to $m_x$
is detected. This can be clearly demonstrated by the following best-fit 
relation to the data:
\begin{equation}
m_{lens}=(1.42\pm0.99)m_v
=(2.23\pm1.15)\beta_{fit}^{-1}m_x,
\end{equation}
in which the uncertainties are the 1$\sigma$ errorbars.  
Meanwhile, Fig.1(a) and (b)
also illustrate the influence of cluster morphologies on the relations
between $m_{lens}$ and $m_v$ and between $m_{lens}$ and $m_x$, 
respectively.
Apparently, it is very unlikely that cluster morphologies can lead 
to a remarkable difference in the results.

\placefigure{fig1}

In Fig.2(a) and (b) we display the variations of $m_{lens}$, $m_v$ and
$m_x$ with the cluster radius. Basically, Fig.2(b) provides a result
similar to the one of Paper I, i.e.,
there is a systematic discrepancy between $m_x$ and $m_{lens}$  inside 
the cluster core radius, and the projected gravitational cluster mass  
obtained with lensing mothed follows a simple power-law of $\sim r^{1.3}$. 
Again, in contrast to $m_x$, $m_v$ agrees statistically
with $m_{lens}$ over all the scales. 

\placefigure{fig2}

Additionally, we have computed the $\beta$ parameter characterizing
the specific energies of the galaxies and the gas in clusters, 
$\beta_{spec}\equiv \sigma^2/(kT/\mu m_p)$ where $\mu m_p=0.59$ is the 
mean particle mass. Our best-fit value with the 18 data points 
(both $T$ and $\sigma$ are available) in Table 1 reads 
$\beta_{spec}=1.29\pm0.71$, while $\beta_{spec}$ reduces to $1.17\pm0.50$ 
if AC114 is excluded. The best-fit relation between the galaxy 
velocity dispersion and the gas temperature is 
\begin{equation}
(\sigma/{\rm km\; s}^{-1})=10^{2.64\pm0.11}(T/{\rm keV})^{0.51\pm0.13}.
\end{equation}
It appears that our best-fit average $\beta_{spec}$ and 
$\sigma$--$T$ relation are consistent with the previous work 
[see Girardi et al. (1996) for summary]. Based on such a good
fitness of eq.(5) alone,  one might conclude that the galaxies 
and the gas 
are in hydrostatic equilibrium with the same cluster potential,
as was claimed by Lubin \& Bahcall (1993). However, our result of
eq.(4) raises a new question of how one could 
reconcile the discrepancy between $m_{lens}$ (or $m_v$) and $m_x$ 
with the good correlation between $\sigma$ and $T$.

Alternatively, it turns out 
from Fig.1(a) that a zero core radius $r_{dc}=0$ for 
the cluster mass profile provides a good fit to the lensing data. 
The best-fit core radius $r_{dc}$ by requiring $m_{v}=m_{lens}$
is $r_{dc}=-0.09\pm0.24$ Mpc, indicative of rather a compact 
dark matter distribution in clusters. 
This result is compatible with the early studies of giant arcs and
statistical lensing of arcs/arclets which report a small 
core radius ($<0.1$ Mpc) for the dark matter profile of the 
arc clusters (e.g. Hammer 1991; Wu \& Hammer 1993; Grossman \& Saha 1994).

While there is a significant evolution of 
X-ray luminosity clusters with redshift (Edge et al. 1990; 
Gioia et al. 1990; Henry et al. 1992), the deficit of the X-ray 
cluster mass may be relevant to the cluster evolution.
Recall that the local $L_x$--$T$ relation 
established at low redshift $z<0.1$ was employed  to estimate the
cluster temperature, whereas most of the clusters in Table 1 are
actually located at intermediate redshift  $z\sim0.2$--$0.5$. Therefore,
it would be useful to examine the dependence of the ratio of $m_{lens}$ to
$m_x$ on the cluster redshift. Our best-fit relation is
\begin{equation}
\frac{m_{lens}}{m_x}=10^{0.09\pm0.10}(1+z)^{1.7\pm0.9}
\end{equation}
Namely, the cluster mass
discrepancy between the X-ray analysis and the gravitational lensing
method is indeed related to the evolutionary history of clusters.
Since $m_x$ is proportional to $T$ according to eq.(1), the
cluster temperature has a similar variation with redshift. This yields 
a temperature ratio of $1.5_{-0.3}^{+0.4}$ 
for cluster at redshift $z=0.33$ relative to the one at $z=0.035$, 
in consistent with the result ($1.4_{-0.3}^{+0.4}$) of Henry et al. (1994). 
Though the cluster temperature
evolution since intermediate redshift is moderate, it 
may account for the mass discrepancy we report in the present paper.

We now discuss briefly the significance of the consistency/discrepancy 
between the cluster masses derived from gravitational lensing, 
dynamical analysis and X-ray observations.

Both gravitational lensing and ``virial'' methods yield nearly the same 
cluster masses, which have several implications: First, 
galaxy velocity dispersion indeed provides a good estimate of cluster
mass. Second, the dark matter particles and the galaxies have approximately
the same velocity dispersion, i.e., there is no velocity bias in
clusters of galaxies. Third, mass follows the light.
These arguments are comparable with the recent dynamical analysis of
the CNOC cluster sample (Carlberg et al. 1996),
which has found strong evidence that
galaxies are effectively in equilibrium with their
host cluster. However, our finding disagrees with the numerical result
that the velocity biasing parameter is $\sim0.7$--$0.8$ (Carlberg \&
Dubinski 1991; Couchman \& Carlberg 1992).

On the other hands, cluster mass estimate based on the X-ray temperature
assuming an isothermal hydrostatic equilibrium has systematically 
underestimated cluster masses by a factor of $\sim2$--$3$ 
as compared with gravitational 
lensing and ``virial'' method, which demonstrates that the gas particles 
may not be a good tracer of the gravitational potential of the cluster. 
The recent high-spectral resolution observations do reveal
the occurrence of complex temperature maps, indicating the  
on-going merger activities in clusters and 
the cluster evolution with cosmic epoch. Overall, 
the simplification of modeling the temperature (isothermal) and  
gas density profile (spherical) is responsible for the deficit of the 
X-ray cluster mass detected in this paper.
One may also attribute
the mass disprepancy to the nonthermal pressure in clusters
(Loeb \& Mao 1994; Ensslin et al. 1997), which
affects the gas particles while produces no effect on 
galaxy distribution and velocity dispersion. 
Finally, it is pointed out that our results still contain large scatters
due to the scarcity of the lensing data, and a large cluster sample will
be needed to confirm our finding.

\acknowledgments

Useful discussion with Fran\c{c}ois Hammer and 
Shude Mao and valuable suggestions by an anonymous referee
are gratefully acknowledged.   
WXP wishes to thank the hospitality of DAEC, Observatoire de 
Paris-Meudon, where part of this research was made. This work 
was supported by the National Science Foundation of China.
 
\clearpage

\begin{table}
\caption{Lensing cluster sample \label{table-1}}
\end{table}

\clearpage
\setcounter{page}{15}

\clearpage

\figcaption{Gravitational lensing derived cluster mass $m_{lens}$
plotted against dynamical cluster mass $m_v$ (a) and X-ray cluster
mass $m_x$ (b). Additionally, cluster morphologies are represented
by different symbols: open circles -- regular, filled circles --
elliptical and asterisks -- irregular. The dashed lines are the 
least square fits of the data to a power-low.
\label{fig1}}

\figcaption{Comparisons of cluster radial mass distributions 
derived from gravitational lensing and dynamical method (a)
and from gravitational lensing and X-ray gas hydrostatic equilibrium (b).
Note that the $\beta$ paramter is taken to be $\beta_{fit}=0.67$ in (b).
\label{fig2}}


\end{document}